\documentclass[preprint,showpacs,preprintnumbers,amsmath,amssymb,endfloats*]{revtex4}
\usepackage{graphicx}

\newcommand{\bes}{\begin{eqnarray}}
\newcommand{\ees}{\end{eqnarray}}

\begin{document}

\thispagestyle{empty}
\title{Experimental test for the conductivity properties
from the Casimir force between metal and semiconductor 
}
\author{F.~Chen, U.~Mohideen}
\affiliation{Department of Physics, University of California,
Riverside, California 92521, USA }
\author{G.~L.~Klimchitskaya}
\affiliation{North-West Technical University, Millionnaya St. 5, St.Petersburg,
191065, Russia}
 \author{V.~M.~Mostepanenko}
\affiliation{Noncommercial Partnership ``Scientific Instruments'', 
Tverskaya St.{\ }11, Moscow, 103905, Russia 
}

\begin{abstract}
The experimental investigation of 
the Casimir force between a large metallized sphere
and semiconductor plate is performed using an atomic
force microscope. Improved calibration and measurement
procedures permitted reduction in the role of different
uncertainties.
Rigorous statistical procedures are applied
for the analysis of random, systematic and total experimental
errors at 95\% confidence. The theoretical Casimir force is computed for
semiconductor plates with different conductivity properties taking
into account all theoretical uncertainties discussed in literature.
The comparison between experiment and theory is done
at both 95 and 70\% confidence. It is demonstrated that the
theoretical results computed for the semiconductor plate used in
experiment are consistent with data. At the same time, theory 
describing a dielectric plate is excluded by experiment at 70\%
confidence. Thus, the Casimir force is proved to be sensitive
to the conductivity properties of semiconductors. 
\end{abstract}

\pacs{12.20.Fv, 12.20.Ds, 68.37.Ps}

\maketitle

\section{Introduction}

The Casimir force \cite{1} is a phenomenon originating from quantum
fluctuations. It depends on both the Planck constant $\hbar$ and
the velocity of light $c$. At the shortest separations of a few
nanometers the relativistic effects become negligible and quantum
fluctuations give rise to the familiar van der Waals force. The
experimental interest in the Casimir effect (see
Refs.~\cite{2,3,4,5,6,7,8,9,10,11,12} and review \cite{13}) is
motivated by the important role played by fluctuating
phenomena in different fields of physics. One of the most
important applications of the Casimir force is in nanotechnology.
Microdevices and microstructures
are fabricated from semiconductors and their characteristic
sizes are of micrometers to
nanometers, i.e., distance scales where the Casimir and van der Waals
forces become dominant \cite{8,13,14,15}. Another notable
application of precision experiments is in the search for new
fundamental interactions and hypothetical particles where the
Casimir force makes the largest contribution to the background 
\cite{11,12,13,16,17,18,19}.

All experiments in Refs.~\cite{2,3,4,5,6,7,8,9,10,11,12} deal with
metallic test bodies. However, the most important
materials used in nanotechnology are
semiconductors with conductivity properties ranging from
metallic to dielectric. Measuring the van der Waals and Casimir
forces between dielectrics has always been a problem due to the
need to eliminate residual charges and contact potential
differences \cite{13}. Semiconductors with
a relatively high conductivity have an advantage that they avoid
accumulation of residual charges
but, at the same time, possess a typical dielectric
dependence of the dielectric permittivity on frequency within
a wide frequency range. This makes it possible to examine the
influence of material properties on the Casimir force and opens new
opportunities to modulate the magnitude and separation dependence
of the force by using semiconductors of different conductivity.
We should note that an early attempt to measure the van der Waals
force in semiconductor surfaces and modify them by light was
reported in Ref.~\cite{19a}. The force was measured between a
glass lens and Si plate and also between the glass lens coated with
Si in a small region and the same Si plate. As the glass lens is
an insulator, the forces due to work function potential differences
could not be controlled. This might explain why no
force change occured on illumination for small separations below
350\,nm \cite{19a} where it should have been most pronounced,
given the approximate inverse third power distance dependence
of the Casimir force for this geometry.

Recent Ref.~\cite{20} presents the first measurement of the Casimir
force between a large gold coated sphere and single crystal
silicon plate with a resistivity $\rho=0.0035\,\Omega\,$cm.
The obtained experimental results were compared with the Lifshitz
theory at zero temperature and excellent agreement was obtained.
It was shown that the Casimir force between Au-Si bodies
decreases with the increase of separation distance more quickly
than between Au-Au bodies. In Ref.~\cite{20} the conclusion was
made that this behavior of a metal-semiconductor system in
comparison with the case of two metals can be used to control the
Casimir force in micro- and nanoelectromechanical systems.
An additional interest in the case of dissimilar materials of the
Casimir plates (metal and dielectric) was stimulated by new
theoretical results. As was shown in Ref.~\cite{21}, the
Casimir pressure and the free energy of a fluctuating field
between metal and dielectric are nonmonotonous functions of
temperature in some temperature intervals. This leads to the
possibility of the negative Casimir entropy although it always
vanishes when the temperature goes to zero (i.e., the Nernst heat
theorem is satisfied).

In this paper we continue the experimental investigation
of the Casimir force between an Au sphere and a Si plate using the
experimental setup described in Ref.~\cite{20}. 
We apply the improved procedures in sample preparation and in
calibration of the setup. This reduces the role of
different uncertainties and leads to important new
results concerning the influence of semiconductor conductivity
on the Casimir force. The analysis of the
obtained experimental data is performed at 95\% confidence, and
random, systematic and total experimental errors are determined
using rigorous methods of mathematical statistics. In particular,
the experimental data are analyzed for the presence of outlying
results. The surface roughness on both the Si plate and the Au
coated sphere is investigated using an atomic force microscope.
This experiment (performed at $T=300\,$K) is shown to be 
insensitive to the temperature corrections predicted by
different approaches to the thermal Casimir force 
(see, e.g., Refs.~\cite{22,23}).
Because of this, the experimental data are compared with the
Lifshitz theory at zero temperature using the tabulated optical
data for the complex refractive indices of Au and Si \cite{24}.
Different kinds of theoretical errors and corrections (in particular,
the correction due to surface roughness) are calculated or estimated.
The resulting total theoretical error is found at 95\% confidence.
The comparison between experiment and theory is performed at both
95\% and 70\% confidence using a rigorous statistical approach.
The theoretical results computed for the p-type B doped Si plate 
of resistivity
$\rho=0.0035\,\Omega\,$cm used in the experiment are found to be 
consistent with the experimental data. At the same time, the
theoretical results computed for high resistivity (``dielectric'')
 Si with $\rho_0=1000\,\Omega\,$cm (as in tables of
Ref.\cite{24}) are excluded experimentally at 70\% confidence.
Hence our experiment demonstrates how the Casimir force between
a metal and a semiconductor is influenced by the conductivity
properties of the latter. Changing the density of free carriers
inside a semiconductor by means of doping or irradiation with
light will lead to changes in the Casimir force.

The paper is organized as follows. In Sec.~II we briefly describe
the experimental setup, and the sample preparation and
characterization including the determination of surface roughness
profiles. Sec.~III describes the measurement of the
deflection coefficient, determination of the
residual electrostatic force and
separation on contact, and calibration.
Sec.~IV contains the measurement results for the
Casimir force and the statistical analysis of the experimental
errors. In Sec.~V the theoretical results for the Casimir force
between a metal and semiconductor in the experimental configuration
are presented, and the theoretical errors are estimated. In Sec.~VI
the comparison between experimental data and theoretical results
computed for Si samples of different resistivity is performed.
Sec.~VII contains our conclusions and discussion.

\section{Experimental setup, and sample preparation and
characterization}

To measure the Casimir force between an Au coated sphere and a
single crystal Si$\langle 100\rangle$ plate, we have used an
improved setup over that in Ref.\cite{6} for both
test bodies coated by Au. In the present experiment the
sphere was coated with an Au layer of 105\,nm thickness and
had a diameter $2R=202.6\pm 0.3\,\mu$m. It was attached to the
cantilever of an atomic force microscope. 
In the atomic force microscopy technique the force is
measured through a deflection of the cantilever attached
to the sphere. The Si plate (doped with B) had
an area $5\times 10\,\mbox{mm}^2$ and thickness of 350\,$\mu$m.
The resistivity of the plate $\rho=0.0035\,\Omega\,$cm was
measured using the four-probe technique. Note that the
resistivities of metals are usually two or three orders of
magnitude lower. Because of this, our Si plate has a relatively
large absorption typical of semiconductors for all frequencies
contributing to the Casimir force at experimental separations.

The main improvements in experimental setup in comparison with
 Ref.\cite{6} are the use of much higher vacuum, and the reduction
of the uncertainty in the determination of absolute sepatations
$z$. A much higher vacuum ($2\times 10^{-7}\,$Torr instead of
$3\times 10^{-2}\,$Torr in Ref.\cite{6}) is needed to maintain
the chemical purity of the Si surface which otherwise oxidizes 
rapidly to SiO${}_2$. The high-vacuum system is needed to
prevent contamination. It consists of oil-free mechanical pumps, 
turbo pumps, and ion pumps. To maintain the lowest pressure during
data acquisition, only the ion pump is used. This helps to reduce
the influence of mechanical noise. The absolute error in the
determination of absolute separations $z$ was reduced to
$\Delta z=0.8\,$nm in comparison with $\Delta z=1\,$nm
in Ref.\cite{6}. This was achieved by using a piezo capable of
traveling a distance of 6\,$\mu$m from initial separation to
contact of the test bodies (in Ref.\cite{6} piezo movement was
used only at separations less than 2\,$\mu$m, and the movement
to larger separations of the plate from the sphere was done
mechanically). 
Such large piezo extensions were also found necessary to
allow time for the decay of noise associated with the
separation of the gold sphere and plate after contact of the
two surfaces. The complete movement of the piezo, $z_{\rm piezo}$,
was calibrated using a fiber optic interferometer \cite{24a}.
To extend and contract the piezo, continuous triangular
voltages at 0.02 Hz are applied to the piezo. Given that the
experiment is done at room temperature, applying of static
voltages will lead to piezo creep and loss of position
sensitivity. The extension and contraction of the piezo were
fit to terms up to fourth order in the applied voltage.
Because of this, the
error in piezo calibration practically does not contribute to
$\Delta z$.

As opposed to Au, the Si surface is very reactive. Because of this,
a special passivation procedure is needed to prepare it to force
measurements. For this purpose nanostrip (a combination of
H${}_2$O${}_2$ and H${}_2$SO${}_4$) is used to clean the surface
of organics and other contaminants. This cleaning, however, 
oxidizes the surface. Then 49\% HF solution was used to etch SiO${}_2$.
This procedure also leads to hydrogen termination of the surface.
The hydrogen termination prevents the re-oxidation of Si surface
as long as it is kept in a high vacuum environment. 
The termination is stable for more than two
weeks under the vacuum conditions described above \cite{26,27}.
We have checked the effectiveness of the passivation technique
to prevent the contamination of the Si surface through the
measurement of the distance dependence of the electrostatic
force resulting from the residual potential difference between
the interacting surfaces (see Sec.~III).

To characterize the topographies of both samples, the Au coating of
the sphere and the surface of Si plate were investigated using an
atomic force microscope. Images resulting from the surface scan
of Au coating demonstrate that the roughness is mostly represented
by the stochastically distributed distortions of about 8--22\,nm.
In Table ~I the fractions $v_k$ of the Au coating with heights
$h_k$ ($k=1,\,2,\ldots ,26$) are presented in columns 3 and 2,
respectively. The surface scan of Si surface shows much smoother 
distortions with the typical heights from 0.4 to 0.6\,nm and with
a maximal height equal to only 1\,nm. The fractions $v_l$ of the
Si surface with heights $h_l$ ($l=1,\,2,\ldots ,11$) are presented
in columns 3 and 2 of Table~II, respectively. Data in Tables~I
and II are
used in Sec.~V to compute the correction to the theoretical
Casimir force due to surface roughness.

\section{Calibration of the setup and determination of the
residual electrostatic force and separation on contact}

All calibrations and determination of the residual electrostatic 
force and of the separation on contact are done immediately
before the Casimir force measurements in the same high vacuum
apparatus. As was already mentioned in Sec.~II, the force is
determined through a deflection of the cantilever. The calibration
of the deflection signal, $S_{\rm def}$, which is 
negative for attractive force and is measured by
using two photodiodes either as a current or a voltage, is done
by applying dc voltages to the Si plate. Care was taken to make
ohmic electrical contacts to the silicon. Direct contact to the
Si plate leads to large residual potentials. Because of this, the
electrical contact was made from a 100\,nm thick gold pad attached 
to the bottom of the plate. The electrical contact to the gold
sphere was accomplished by applying a very thin gold coating
to the cantilever. In addition, a small correction has to be
applied to the separation distance between the gold sphere and
the Si plate due to the movement of the cantilever. The actual
separation distance $z$ between the bottom of the gold sphere and 
the Si plate is given by 
\begin{equation}
z=z_{\rm piezo}+S_{\rm def}\,m+z_0,
\label{eq0}
\end{equation}
\noindent
where $m$ is the deflection coefficient in units of nm per
unit deflection signal.

The measurement of $m$ was performed
by applying
different dc voltages $V$ between $+0.2$ to $-0.4\,$V 
to the plate. To find the coefficient $m$, the cantilever 
deflection signal was
measured as a function of the distance. The 0.02\,Hz
triangular wave was applied to the piezo to change the
distance between the sphere and the plate. 
Larger applied voltages lead to more cantilever deflection
and therefore earlier contact of the two surfaces. The change
in the contact position of the sphere and the plate as 
a function of the applied voltage can then be used to
measure deflection coefficient $m$ \cite{6}.
In order to determine the contact of the two surfaces
precisely, 32768 data points at equal time intervals were
acquired for each force measurement. In distinction to
Ref.~\cite{6}, in cases, where the contact was
between two neighboring data points, a linear interpolation
was used to identify the exact value. The deflection
coefficient was found to be $m=43.3\pm 0.3\,$nm per unit
deflection signal. This value was used to correct the
separation distance in all measurements.

The determination of the residual potential 
difference between the two
surfaces $V_0$ was
performed by the fit between electric force measurements far 
away from contact (where the Casimir force is practically zero)
and exact force-distance relation. To measure the force, the
calibration of the deflection signal was performed. In this work
an improved method, rather than simple application of a dc 
voltage $V$ to the plate, was used. This was done to avoid
systematic errors due to scattered laser light. In addition to
the application of the dc voltage $V$ to the Si plate, 
square voltage pulses of amplitudes in the range $\pm 0.4\,$V 
and time interval corresponding to a 
separation distance between 1 to 5\,$\mu$m were also applied
to the plate. 
The dc voltage was close to the residual potential difference
$V_0$ in order to decrease systematic errors due to large
deflections.
Fig.~1 shows the deflection signal of the
cantilever in response to both the applied dc voltage and
the square pulse as a function of the separation distance
between the gold sphere and Si plate. By measuring only
the difference in signal during the pulse allows one to
avoid the need for a background subtraction. Also the
large width of the pulse allowed checks for the distance
dependence of the residual potential and any position
dependence in the calibration.

The average values
of the measured electric forces as a function of separation were
used to fit the exact force-distance relation \cite{25}
\begin{equation}
F(z)=2\pi\varepsilon_0(V-V_0)^2
\sum\limits_{n=1}^{\infty}
\frac{\coth\alpha-n\coth n\alpha}{\sinh n\alpha},
\label{eq1}
\end{equation}
\noindent
where $\cosh\alpha=1+z/R$, $V$ is the voltage applied to the
Si plate, and $\varepsilon_0$ is a permittivity of
a vacuum. Nonzero value of $z$ at contact, $z_0$, is due to
surface roughness. This is a distance between the zero roughness
levels [see Eq.~(\ref{eq22}) below]. For application to this
experiment, Eq.~(\ref{eq1}) is conveniently rearranged as follows
\begin{equation}
F(z)=-2\pi\varepsilon_0(V-V_0)^2
\sum\limits_{i=-1}^{6}
c_i
\left(\frac{z}{R}\right)^i\equiv X(z)(V-V_0)^2,
\label{eq2}
\end{equation}
\noindent
where
\begin{eqnarray}
&&
c_{-1}=0.5,\quad c_0=-1.18260,\quad c_1=22.2375,\quad
c_2=-571.366,
\nonumber \\
&&
c_3=9592.45,\quad c_4=-90200.5,\quad c_5=383084,
\quad c_6=-300357.
\nonumber
\end{eqnarray}
\noindent
Within the separation region from 1.8 to 5\,$\mu$m, the relative
error introduced by the use of Eq.~(\ref{eq2}) instead of
Eq.~(\ref{eq1}) does not exceed $1.5\times 10^{-5}$. 
Eq.~(\ref{eq2}) at fixed separation $z$ is used to fit the
difference signal and the residual potential difference was
determined to be $V_0=-0.114\pm 0.002\,$V. The calibration of
the deflection signal was also performed using the same
procedure. The force calibration constant was determined
to be $1.440\pm 0.007\,$nN per unit cantilever deflection
signal.

The value of $V_0$ was found to be independent of separation.
This confirms the absence of localized charges because they would
lead to dipole and other multipolar electrostatic fields, resulting
in a residual potential difference varying with distance. As was
mentioned above, the relatively high conductivity of the Si plate,
used in this experiment, is important in preventing the formation
of the localized charges. The independence of $V_0$ on separation
confirms also the absence of any contamination of the Si surface.

The separation distance on contact of the two surfaces $z_0$ needs
to be independently determined for a comparison of the measured
Casimir force to the theory. To achieve this goal, we apply
different dc voltages to the Si plate (like it was done in the
measurement of $m$) and measure the electrostatic
force as a function of separation. This
measurement at each voltage was repeated
5 times and the average signal curve was obtained.
A compensation dc voltage equal to $V_0$ was applied to the
plate and the resulting deflection signal was subtracted from
the signal corresponding to electrostatic force curves at all
other dc voltages. This procedure eliminates the need for
subtraction of the background and Casimir forces from the
electrostatic force curves.
In difference from Ref.~\cite{6}, in the
determination of $z_0$ we attempted to reduce the role of
uncertainties in $V_0$. The procedure used here also gives one more
way for the determination of $V_0$ and check of its distance
independence. It is as follows: At a fixed separation $z$
different voltages $V$ are applied to the plate and the
electrostatic force is plotted as a function of $V$. The parabolic
dependence of this force [see Eq.~(\ref{eq2})] is used to determine
the values of $V_0$ and $X(z)$ \cite{Overbeek}. 
This is repeated for many
different $z$. The value of $V_0$ was found to be the same as that
determined earlier and to be independent on separation
distance. Note that this determination is also independent of
errors in the cantilever calibration. In order to determine
$z_0$, $X(z)$ is then plotted as a function of $z$ and fitted to
Eq.~(\ref{eq2}). The value of $z_0$ so determined is 32.1\,nm. 
The uncertainty in the quantity $z_0+S_{\rm def}\,m$ 
[see Eq.~(\ref{eq0})] due to both the uncertainty in $m$
and calibration was found to be
0.8\,nm. As was mentioned in Sec.~II, the error in piezo
calibration contributes negligibly to the error in measurement
of absolute separations $\Delta z$. Because of
this, with account of Eq.~(\ref{eq0}), we arrive at
$\Delta z=0.8\,$nm. (Note that in Ref.~\cite{20} for the sake of
brevity the uncertainty in $z_0+S_{\rm def}\,m$ was attributed to
$z_0$.)

\section{Measurement results and experimental errors}

The Casimir force between the sphere and the plate was measured as 
a function of distance. In doing so the sphere was kept grounded
while a compensating voltage $V_0$ was applied to the plate to
cancel the residual electrostatic force. The distance was varied 
continuously from large to short separations by applying continuous 
triangular voltages at 0.02\,Hz to the piezo.
The piezo was extended to its maximum range over 6\,$\mu$m. 
The force data $F^{\rm expt}(z_i)$ were
collected at 32768 equal time intervals as the distance between the
sphere and plate was changed. This measurement was repeated
for $n=65$ times.
A great advantage of the atomic force microscopy technique in the
averaging is that  the contact point between the two surfaces
$z_0$ provides a starting point for alignment of all the 65
measurements. Nevertheless, thermal noise in the cantilever
deflection signal, $S_{\rm def}$, leads to noise in the
corresponding separations $z$. To account for this in the
averaging, the separation distance is divided into a grid of
32768 equidistant points separated by 0.17\,nm. For each measured
Casimir force-distance curve, the value of the force at the
grid point is computed using linear interpolation of the
neighboring two data points. Because the separation distance
between neighboring points is small as 0.17\,nm, higher order
interpolation procedures were not required. Also the noise
spectrum and amplitude of the interpolated data were confirmed
to be the same as the raw data. This allowed the averaging of 
the 65 Casimir force mesurements even including the effect of 
the change in the separation distance due to the thermal noise
of a cantilever.

Below we present the measurement results within the 
separation range
from 62.33\,nm to 349.97\,nm. This distance range (containing
1693 points) is chosen for the reason that for larger separations
the experimental relative error of force measurements 
caused by the noise exceeds
100\% (see below), i.e., the data are not informative.
Data for separations below 62.33\,nm up to contact are not
presented as nonlinearities associated with the ``jump to
contact'' introduce uncontrollable errors into the force
measurement.

For convenience we
denote the separations by $z_{ij}$, where $1\leq i\leq 1693$ and
$1\leq j\leq 65$. In the present experiment the separations
$z_{ij}$ with fixed $i$ are the same in all measurement sets, i.e.,
do not depend on $j$. 
The mean values of the force
\begin{equation}
{\bar{F}}_i\equiv {\bar{F}}^{\rm expt}(z_i)=\frac{1}{n}
\sum\limits_{j=1}^{n}F^{\rm expt}(z_{ij})
\label{eq3}
\end{equation}
\noindent
are plotted in Fig.~2 as a function of separation within  
the chosen separation range. As is seen
from Fig.~2, at short separations
the mean force ${\bar{F}}^{\rm expt}(z_i)$ is uniform,
i.e., changes smoothly with the change of $z_i$.
Several values of the mean experimental force are listed in column 
2 of Table~III.

Before proceeding with the analysis of random errors, we examine
the experimental data for the presence of so-called ``outlying''
results. For this purpose it is necessary to consider the quantity
\cite{28,29}
\begin{equation}
T_i=\frac{1}{s_{{F}_i}}\max|F^{\rm expt}(z_{ij})-
{\bar{F}}_i|,
\label{eq4}
\end{equation}
\noindent
where maximum is taken with respect to $j$, and the variance of the
force is
\begin{equation}
{s_{{F}_i}}=\left\{\frac{1}{n-1}
\sum\limits_{j=1}^{n}\left[F^{\rm expt}(z_{ij})-
{\bar{F}}_i\right]^2\right\}^{1/2}.
\label{eq5}
\end{equation}
\noindent
According to standard statistical procedure, the measurement set $j$
contains the outlying result at a point $z_i$ with
a confidence probability $\beta$ if the inequality
$T_i>T_{n,1-\beta}$ is satisfied, where $T_{n,1-\beta}$ are
tabulated quantities \cite{28,29}. Usually the set of measurements
is rejected if it contains results which are outlying at a
confidence probability 90\%. Using this statistical criterion we
have analyzed all $n=65$ sets of measurements and found that $T_i$
changes from 2.4 to 2.7. For a bilateral check [which is applicable
because in our case the deviations of $F^{\rm expt}(z_{ij})$ from
${\bar{F}}_i$ are both positive and negative] $T_{65,0.1}=3.2$
\cite{28,29}. Hence it follows that there are no outlying results 
among our measurement data and all of them should be used in the
determination of the experimental errors.

We start with random errors and find the variance of the mean force
\begin{equation}
{s_{\bar{F}_i}}=\left\{\frac{1}{n(n-1)}
\sum\limits_{j=1}^{n}\left[F^{\rm expt}(z_{ij})-
{\bar{F}}_i\right]^2\right\}^{1/2}.
\label{eq6}
\end{equation}
\noindent
Direct calculation shows that the variance of the mean, 
${s_{\bar{F}_i}}$, is not uniform, i.e., changes stochastically
in going from one $z_i$ to another. The computational results
for this variance as a function of separation are shown in
Fig.~3. To smooth the variance of the mean, we have used a special
procedure developed in mathematical statistics \cite{30,31}. For each
separation $z_i$ we consider $N$ neighboring points (half of them from
the left and half from the right of $z_i$; $N=30$ in this experiment).
Then the smoothed variance of the mean force at a point $z_i$ is
given by \cite{30,31}
\begin{equation}
{\tilde{s}_{\bar{F}_i}}=\left[N\sum\limits_{k=1}^{N}
\lambda_k^2{s}_{\bar{F}_i}^2\right]^{1/2},
\label{eq7}
\end{equation}
\noindent
where $\lambda_k$ are the statistical weights. The two different
sets of the statistical weights can be chosen:
\begin{equation}
\lambda_k^{(1)}=\frac{1}{N},\quad
\lambda_k^{(2)}=\frac{1}{c_k\sum\limits_{i=1}^{N}c_i^{-1}},
\label{eq8}
\end{equation}
\noindent
where constants $c_i$ are determined from
\begin{equation}
{s}_{\bar{F}_1}^2:{s}_{\bar{F}_2}^2:\ldots :{s}_{\bar{F}_N}^2=
c_1:c_2:\ldots :c_N.
\nonumber
\end{equation}
\noindent
Here we choose $\lambda_k=\lambda_k^{(1)}$ in Eq.~(\ref{eq7}).
This leads to larger values of ${\tilde{s}_{\bar{F}_i}}$ and, thus,
is a more conservative estimate, overestimating the random errors.
In our experiment the application of this statistical procedure
results in approximately the same variance for all $z_i$
equal to 
${\tilde{s}_{\bar{F}_i}}\equiv {\tilde{s}_{\bar{F}}}\approx 1.5\,$pN.

Now we are in a position to find the absolute random error for the
Casimir force measurements in the configuration of Au sphere and Si
plate. Using the Student's $t$-distribution with a number of degrees
of freedom $f=n-1=64$ and choosing the 95\% confidence probability
($\beta=0.95$), we obtain $p=(1+\beta)/2=0.975$ and $t_p(f)=2$
\cite{32}. Thus, the absolute random error does not depend on
separation and is given by
\begin{equation}
\Delta^{\!\rm rand}F^{\rm expt}={\tilde{s}_{\bar{F}}}t_p(f)=
3.0\,\mbox{pN}.
\label{eq9}
\end{equation}
\noindent
The relative random error is defined as
\begin{equation}
\delta^{\rm rand}F^{\rm expt}(z_i)=
\frac{\Delta^{\!\rm rand}F^{\rm expt}}{|{\bar{F}}^{\rm expt}(z_i)|}.
\label{eq10}
\end{equation}
\noindent
The values of the relative random error as a function of separation
are given in the second column of Table~IV. It achieves the smallest
value equal to 0.78\% at the shortest separation $z=62.33\,$nm
and increases with the increase of separation.

There are following four systematic errors in this experiment
\cite{20}: $\Delta_1^{\!{\rm syst}}F^{\rm expt}\approx 0.82\,$pN
due to the error in force calibration;
$\Delta_2^{\!{\rm syst}}F^{\rm expt}\approx 0.55\,$pN due to
noise when the calibration voltage is applied to the cantilever;
$\Delta_3^{\!{\rm syst}}F^{\rm expt}\approx 0.31\,$pN due to the 
instrumental sensitivity; and
$\Delta_4^{\!{\rm syst}}F^{\rm expt}\approx 0.12\,$pN due to
the restrictions on computer resolution of data. Systematic errors
are random quantities characterized by a uniform distribution.
The best estimate for their combination, i.e., for a total
systematic error at a given confidence probability $\beta$, is
given by \cite{28}
\begin{equation}
\Delta^{\!{\rm syst}}F^{\rm expt}=\min\left[\sum\limits_{i=1}^{J}
\Delta_i^{\!{\rm syst}}F^{\rm expt},\,
k_{\beta}^{(J)}\sqrt{\sum\limits_{i=1}^{J}
\left(\Delta_i^{\!{\rm syst}}F^{\rm expt}\right)^2}\right],
\label{eq11}
\end{equation}
\noindent
where in our experiment $J=4$, and $k_{\beta}^{(J)}$ is a tabulated
coefficient. Using the value $k_{0.95}^{(4)}=1.12$ we obtain from
Eq.~(\ref{eq11}) $\Delta^{\!{\rm syst}}F^{\rm expt}=1.17\,$pN at
95\% confidence. In the third column of Table~IV the values of the
relative systematic error
\begin{equation}
\delta^{\rm syst}F^{\rm expt}(z_i)=
\frac{\Delta^{\!\rm syst}F^{\rm expt}}{|{\bar{F}}^{\rm expt}(z_i)|}
\label{eq12}
\end{equation}
\noindent
are listed at different separations. As is seen from this Table,
the smallest relative systematic error of 0.31\% is achieved at
the shortest separation. Comparing second and third columns in Table~IV, 
we conclude that at all separations the magnitude of the systematic
error comprises about 0.4 of the random error.

To find the total experimental error in the Casimir force
measurements, one should combine the  random and
systematic errors obtained above. 
They are described by a normal (or Student)
distribution, and by a combination of uniform distributions,
respectively. To be very conservative, we assume that the
combination of systematic errors is also described by a uniform
distribution (other assumptions would lead to a smaller total
error). Different methods for combining random and systematic
errors are described in literature \cite{28}. Here we use the rule
based on the consideration of the ratio
\begin{equation}
r(z_i)=
\frac{\Delta^{\!{\rm syst}}F^{\rm expt}(z_i)}{\tilde{s}_{\bar{F}_i}}.
\label{eq12a}
\end{equation}
\noindent
If the inequality $r<0.8$ is satisfied, the systematic error in
most cases can be neglected and the total error at 95\% confidence
is given by
\begin{equation}
\Delta^{\!{\rm tot}}F^{\rm expt}(z_i)=
\Delta^{\!{\rm rand}}F^{\rm expt}(z_i).
\label{eq12b}
\end{equation}
\noindent
If $r>8$, the random error can be neglected and the total error
at the same confidence is
\begin{equation}
\Delta^{\!{\rm tot}}F^{\rm expt}(z_i)=
\Delta^{\!{\rm syst}}F^{\rm expt}(z_i).
\label{eq12c}
\end{equation}
\noindent
In the region where $0.8\leq r\leq 8$ it is recommended \cite{28}
to use the expression
\begin{equation}
\Delta^{\!{\rm tot}}F^{\rm expt}=
q_{\beta}\left[\Delta^{\!{\rm rand}}F^{\rm expt}+
\Delta^{\!{\rm syst}}F^{\rm expt}\right]
\label{eq13}
\end{equation}
\noindent
with a maximum possible value of the coefficient $q_{0.95}=0.8$
determined at 95\% confidence. 
In our case the random error exceeds the systematic one, and
the value of $r$ in Eq.~(\ref{eq12a}) is constant and close
to 0.8. Because of this, to be conservative, we choose the
largest of the total errors given by Eqs.~(\ref{eq12b}) and
(\ref{eq13}), i.e.,
$\Delta^{\!\rm tot}F^{\rm expt}\approx 3.33\,$pN.

In the fourth column of Table~IV we present the values of the
relative total experimental error
\begin{equation}
\delta^{{\rm tot}}F^{\rm expt}(z_i)=
\frac{\Delta^{\!{\rm tot}}F^{\rm expt}}{|{\bar{F}}^{\rm expt}(z_i)|}
\label{eq14}
\end{equation}
\noindent
at different separations. It is equal to only 0.87\% at the shortest 
separation and achieves 64\% at a separation $z=299.99\,$nm.
At $z=350\,$nm the total experimental error
exceeds 100\%.
Comparing second, third and fourth columns in Table~IV, one can
conclude that at all separations the major contribution to the
total experimental error is given by the random error.

\section{Theoretical approach to the Casimir force between
metal and semiconductor}

As was demonstrated in the preceeding section, the lowest total 
experimental errors ranging from 0.87 to 5.3\% are achieved at
separations $z\leq 120\,$nm. At such short separations the thermal
effects are not important (see below for the magnitudes of
predicted thermal corrections) and one can use the Lifshitz formula
at zero temperature for the Casimir (van der Waals) force acting
between a Si plate and an Au sphere
\begin{eqnarray}
&&
F(z)=\frac{\hbar R}{2\pi}\int_{0}^{\infty}k_{\bot}dk_{\bot}
\int_{0}^{\infty}d\xi\left\{\ln\left[1-
r_{\|}^{(1)}(\xi,k_{\bot})r_{\|}^{(2)}(\xi,k_{\bot})
e^{-2zq}\right]\right.
\nonumber \\
&&\phantom{aaaa}
+\left.\ln\left[1-
r_{\bot}^{(1)}(\xi,k_{\bot})r_{\bot}^{(2)}(\xi,k_{\bot})
e^{-2zq}\right]\right\}.
\label{eq15}
\end{eqnarray}
\noindent
Here the reflection coefficients for two independent polarizations
of the electromagnetic field are
\begin{equation}
r_{\|}^{(p)}(\xi,k_{\bot})=
\frac{\varepsilon^{(p)}(i\xi)q-k^{(p)}}{\varepsilon^{(p)}(i\xi)q+
k^{(p)}}, \quad
r_{\bot}^{(p)}(\xi,k_{\bot})=\frac{k^{(p)}-q}{k^{(p)}+q},
\label{eq16}
\end{equation}
\noindent
where
\begin{equation}
q^2=k_{\bot}^2+\frac{\xi^2}{c^2},\quad
{k^{(p)}}^2=k_{\bot}^2+\varepsilon^{(p)}(i\xi)\frac{\xi^2}{c^2}.
\label{eq17}
\end{equation}
\noindent
$\varepsilon^{(p)}(\omega)$ is the dielectric permittivity of Au
($p=1$) and Si ($p=2$), and $k_{\bot}$ is the magnitude of the
wave vector projection on the Si plate.
Note that in Eq.~(\ref{eq15}) we have replaced a 105\,nm Au layer
for an Au semispace. Using the Lifshitz formula for layered structures
[13], it is easy to calculate the force error due to this
replacement. At a typical separation of 100\,nm it is only 0.0095\%,
thus justifying the use of the semispace.

The dielectric permittivities of Au and Si along the imaginary
frequency axis are computed by means of the dispersion relation
\begin{equation}
\varepsilon^{(p)}(i\xi)=1+\frac{2}{\pi}
\int_{0}^{\infty}d\omega
\frac{\omega\mbox{Im}\varepsilon^{(p)}(\omega)}{\omega^2+\xi^2},
\label{eq18}
\end{equation}
\noindent
where $\mbox{Im}\varepsilon^{(p)}(\omega)$ are calculated using the
tabulated optical data for the complex index of refraction \cite{24}.
For Au the data for $\omega\geq 0.125\,$eV are available
($1\,\mbox{eV}=1.519\times 10^{15}\,$rad/s). To obtain
$\varepsilon^{(1)}(i\xi)$ by Eq.~(\ref{eq18}) with sufficient
precision one should extrapolate the tabulated data to the region
of smaller frequencies. This is usually done (see, e.g., 
Ref.~\cite{33}) with the help of the imaginary part of the Drude
dielectric function
\begin{equation}
\mbox{Im}\varepsilon^{(1)}(\omega)=
\frac{{\omega_p^{(1)}}^2\gamma^{(1)}}{\omega\left(\omega^2+
{\gamma^{(1)}}^2\right)},
\label{eq19}
\end{equation}
\noindent
where $\omega_p^{(1)}=9.0\,$eV, and $\gamma^{(1)}=0.035\,$eV are
the plasma frequency and relaxation parameter of Au, respectively.
The most precise computational results for $\varepsilon^{(1)}(i\xi)$
using this procedure are presented in Ref.~\cite{33}.

For dielectric Si with a resistivity $\rho_0=1000\,\Omega\,$cm the 
tabulated data for the complex refractive index are available for
$\omega>0.00496\,$eV \cite{24}. Thus, there is no need
for any extrapolation of data to smaller frequencies. The computational
result for the dielectric permittivity of dielectric Si along the
imaginary frequency axis, $\tilde{\varepsilon}^{(2)}(i\xi)$, was
obtained in Ref.~\cite{33}. It is shown by the dashed line in Fig.~4.
This line can be extended to zero frequency leading to the value
of the static dielectric permittivity of Si
${\tilde{\varepsilon}}^{(2)}(0)=11.67$. The values of
${\tilde{\varepsilon}}^{(2)}(i\xi)$ can be used in precise computations 
of the Casimir and van der Waals interactions between the test bodies
made of dielectric Si. Note that the use of the analytical
approximation for ${\tilde{\varepsilon}}^{(2)}(i\xi)$, suggested in
Ref.~\cite{34}, leads to about 10\% error in the magnitudes of the
Casimir force, and, thus, is not suitable for the comparison
with precise measurements.

In our experiment, however, the Si plate of much lower resistivity
$\rho=0.0035\,\Omega\,$cm than in tables is used. 
This resistivity corresponds to B doped Si. 
The plasma frequency
for such Si is equal to 
\begin{equation}
\omega_p^{(2)}=\frac{e\sqrt{n}}{\sqrt{\varepsilon_0m^{\ast}}}
\approx 7\times 10^{14}\,\mbox{rad/s}.
\label{20}
\end{equation}
\noindent
Here the doping concentration leads to a carrier density 
$n\approx (2.9-3.2)\times 10^{19}\mbox{cm}^{-3}$.
This value of $n$ corresponding to the sample of 
resistivity $\rho=0.0035\,\Omega\,$cm 
was obtained from Fig.~2.18 in Ref.~\cite{34a}. The
optical effective mass for B doped Si used in this experiment is 
$m^{\ast}=0.206m_e$ \cite{34b}. The respective relaxation parameter
of the Drude model
$\gamma^{(2)}$ is determined from
$\gamma^{(2)}=\varepsilon_0\rho\,{\omega_p^{(2)}}^2\approx 1.5\times
10^{14}\,$rad/s.
Since the optical properties of Si at the frequencies 
making a nonnegligible contribution
to the Casimir force depend on the concentration of
charge carriers, the optical data of dielectric Si should be
adapted for our case. This is achieved \cite{24} by adding
the imaginary part of the Drude dielectric function to the
imaginary part of the dielectric permittivity obtained 
from tables. For our Si plate of lower resistivity $\rho$ this results
in
\begin{equation}
\varepsilon^{(2)}(i\xi)={\tilde{\varepsilon}}^{(2)}(i\xi)+
\frac{{\omega_p^{(2)}}^2}{\xi\left(\xi+\gamma^{(2)}\right)}.
\label{eq21}
\end{equation}
\noindent
In Fig.~4 the dependence of $\varepsilon^{(2)}$ on frequency is shown
by the solid line. Once the dielectric permittivities of Au and Si
along the imaginary frequency axis have been computed, the Casimir
force can be found by Eqs.~(\ref{eq15})--(\ref{eq17}).

For comparison of theory with experiment, the theoretical results
obtained using the Lifshitz formula should be corrected for the
presence of surface roughness \cite{6,13}. The topography data
in Tables~I and II of Sec.~II allow one to determine the zero roughness
levels $H_0^{(p)}$ relative to which the mean values of the
functions, describing roughness, are zero:
\begin{equation}
\sum\limits_{k=1}^{26}(H_0^{(1)}-h_k)v_k=
\sum\limits_{l=1}^{11}(H_0^{(2)}-h_l)v_l=0.
\label{eq22}
\end{equation}
\noindent
Calculations using Tables~I and II lead to $H_0^{(1)}=15.352\,$nm,
$H_0^{(2)}=0.545\,$nm. As was mentioned in Sec.~III, all absolute
separations between the test bodies are measured between the
zero roughness levels determined in Eq.~(\ref{eq22}).
 This permits one to achieve rather
low absolute error in separation measurements ($\Delta z=0.8\,$nm
in this experiment) in spite of the fact that the roughness
heights $h_k$ are much larger (at least on the sphere).

In the framework of the additive approach \cite{13} the Casimir
force corrected for the presence of roughness can be calculated
as
\begin{equation}
F^{\rm theor}(z_i)=\sum\limits_{k=1}^{26}
\sum\limits_{l=1}^{11}v_kv_lF(z_i+H_0^{(1)}+H_0^{(2)}-h_k-h_l),
\label{eq23}
\end{equation}
\noindent
where $F$ is given by the Lifshitz formula (\ref{eq15}). In this
experiment the influence of surface roughness is very moderate. 
For example, if separation increases from 62.33 to 100.07\,nm,
the ratio $F^{\rm theor}/F$ computed using Eqs.~(\ref{eq15}) and
(\ref{eq23}) decreases from 1.015 to 1.006. Thus, the contribution
of surface roughness achieves the maximum value of 1.5\% at the
shortest separation and decreases to only 0.6\% at $z=100.07\,$nm.
Note that Eq.~(\ref{eq23}) takes into account the nonmultiplicative
effects in corrections due to surface roughness and to realistic
conductivity properties. A more simple multiplicative approach 
describes the surface roughness by the stochastic functions and
presents the theoretical force in the form \cite{13}
\begin{equation}
F_m^{\rm theor}(z_i)=F(z_i)\left\{1+6\left[
\left(\frac{\delta_{st}^{(1)}}{z_i}\right)^2+
\left(\frac{\delta_{st}^{(2)}}{z_i}\right)^2\right]\right\},
\label{eq24}
\end{equation}
\noindent
where the variances of the random processes describing the
stochastic roughness are given by
\begin{equation}
\delta_{st}^{(1)}=\left[
\sum\limits_{k=1}^{26}(H_0^{(1)}-h_k)^2v_k\right]^{1/2}=
3.446\,\mbox{nm}, \quad
\delta_{st}^{(2)}=\left[
\sum\limits_{l=1}^{11}(H_0^{(2)}-h_l)^2v_l\right]^{1/2}=
0.111\,\mbox{nm}.
\label{eq25}
\end{equation}
\noindent
Calculations using Eq.~(\ref{eq24}) show that the ratio
$F_m^{\rm theor}/F$ changes from 1.019 to 1.007 when separation 
increases from 62.33 to 100.07\,nm. Thus, the maximal error of
the multiplication approach in comparison with Eq.~(\ref{eq23})
is only 0.4\%. At $z=100.07\,$nm the error of the multiplication
approach decreases to 0.1\% of the force.

It is well known that the additive approach is the approximative
one, and there are corrections to Eq.~(\ref{eq23}) due to the
diffraction-type and correlation
effects \cite{35,36}. For the case of Au-Au
test bodies these corrections were estimated in Ref.~\cite{6}
using the results of Ref.~\cite{35}. In our case of Au and Si
test bodies the upper limit for the ratio $F_m^{\rm theor}/F$
computed within the multiplicative approach with account of
the contribution of diffraction and correlation
effects is 1.0204 at the
shortest separation. This should be compared with 1.019
(where these effects were neglected) leading to
only 0.14\% difference. It is easily seen that the contribution
of diffraction and correlation effects remains negligible over 
the separation
range where the present experiment achieves high precision. 
At larger separations, however, where the contribution of
these effects in the roughness correction
increases, the complete roughness correction
becomes negligible and the precision of the measurement decreases
(see the second paper in Ref.~[6] for more details).

Now we discuss the theoretical errors which may occur in the
computation of the Casimir force using Eqs.~(\ref{eq15}) and
(\ref{eq23}). The first error $\delta_1F^{\rm theor}$ arises from
the variations of the tabulated optical data for the complex index 
of refraction for both Au and dielectric Si
(recall that we have taken these data from tables in
Ref.~\cite{24}). As was shown in Ref.~\cite{6}, this error is no 
larger than 0.5\%. Being conservative, we set 
$\delta_1F^{\rm theor}=0.5$\% at all separations. 
In addition the tabulated optical data of the
dielectric Si has to be modified due to the use of B doped Si
according to Eq.~(25).
Note that the
calculation results for the Casimir force are rather stable to the
change of the Drude model parameters. Thus, a change of
$\omega_p^{(2)}$ even by a factor of 1.5 leads to less than a 1\% change 
in the Casimir force within the separation region from 62 to 150\,nm.
There may occur rare Au samples of bad quality leading to up to 2\%
deviations in the Casimir force at short separations in comparison
with that computed using the tables. If such sample were used in the
experiment, the theoretical Casimir force would deviate from
the experimental data. Such deviations of the theory from
experiment must be considered not as an error (note that they
can only diminish the magnitude of the Casimir force) but as a
correction similar to the roughness correction. 
The verification of the hypothesis of the
presence of such types of corrections can be easily performed with
the help of standard statistical procedures.

A second theoretical error is caused by the use of the proximity
force theorem in Eq.~(\ref{eq15}). The exact value for the Casimir
force in the configuration of a sphere above a plate is still unknown.
Its magnitude may be both smaller or larger with equal probability 
than the estimate given
by the proximity force theorem. The upper
limit of the error introduced by the use of the proximity force
theorem was estimated as $\delta_2F^{\rm theor}=z/R$ \cite{37,38}.

Both errors $\delta_1F^{\rm theor}$ and $\delta_2F^{\rm theor}$ 
are described by a uniform distribution and, thus, are analogous to
systematic errors. They can be combined by using Eq.~(\ref{eq11})
with $J=2$, $k_{0.95}^{(2)}=1.10$ which is applicable to both
absolute and relative errors. The values of the obtained theoretical
error, $\delta_0F^{\rm theor}$, are presented in the fifth column
of Table~IV at different separations. This combined error slowly
increases with the increase of separation and takes values between
0.55 and 0.64\%.

In addition to the  theoretical errors
$\delta_1F^{\rm theor}$ and $\delta_2F^{\rm theor}$
discussed above,  there exist
the uncertainties due to patch potentials, nonlocal effects and
that due to finite size of the plate which are not taken into account
in Eqs.~(\ref{eq15}) and (\ref{eq23}). All these uncertainties are
shown to be negligibly small \cite{6,12}, and, thus, 
are neglected in the calculation of theoretical errors.

There is one more effect which is not taken into account in full in
Eq.~(\ref{eq15}) but merits consideration. The measurements were
performed at $T=300\,$K. However, Eq.~(\ref{eq15}) is written in 
the form of an integral over the imaginary frequency which is related 
to zero temperature. The problem of the thermal Casimir force between
conducting materials has been a matter of debate over the last few years
and two different approaches to its resolution were suggested (see,
e.g., Refs.~\cite{22,23}). According to Ref.~\cite{22}, the
thermal corrections at short separations are negligibly small for
any conducting materials in qualitative agreement with the case
of ideal metals \cite{13}. On the contrary, Ref.~\cite{23} predicts
thermal correction which can be several hundreds 
times greater than that for
ideal metals at short separations. In our experiment, however, even
the large thermal correction predicted in Ref.~\cite{23} is
negligibly small. For example, for Au sphere and Si plate of
resistivity $\rho$ at $z=60\,$nm the approach of Ref.~\cite{23}
leads to the thermal correction equal to 0.015\% of the
zero-temperature force. This correction increases up to 0.039\%
at $z=90\,$nm and to 0.13\% at $z=120\,$nm. With increasing
separation, however, the precision of the measurements decreases
more rapidly, so that the predicted thermal correction remains
negligible. For Au sphere above a plate made of dielectric Si,
the thermal correction is even smaller. Thus, at $z=100\,$nm
it is equal to only 0.005\% of the zero-temperature force.
It should also be noted  that the optical properties of
the materials are determined not at zero but at room temperature. 
Because of this, even the Lifshitz formula (\ref{eq15}) includes
some dependence of temperature as a parameter.
Thus the use of the Lifshitz formula at zero 
temperature for the comparison with experimental data is well 
justified.

In the foregoing we did not discuss one more error which is in fact
not theoretical but plays some role in the comparison of theory with
experiment. This arises from the fact that the theoretical force
should be calculated in the experimental points $z_i$ which are
determined with an error $\Delta z$ \cite{39}. Bearing in mind that 
the leading theoretical dependence has the form $R/z^3$, this leads
to the error $\delta_3F^{\rm theor}=0.95(\Delta R/R+3\Delta z/z$)
determined at 95\% confidence.
The values of $\delta_3F^{\rm theor}$ at different separations are
presented in column 6 of Table~IV. It is seen that at shortest
separation $\delta_3F^{\rm theor}=3.8\,$\%, i.e., much larger than
the proper theoretical error $\delta_0F^{\rm theor}$. At $z=299.99\,$nm
$\delta_3F^{\rm theor}=0.9\,$\%, i.e., also larger than
$\delta_0F^{\rm theor}$. In a conservative way the errors 
$\delta_0F^{\rm theor}$ and $\delta_3F^{\rm theor}$ can be combined
using the analogues of Eqs.~(\ref{eq12c}) and (\ref{eq13})
written in terms of relative errors. 
Thus, we obtain what we call
the total theoretical error $\delta^{\rm tot}F^{\rm theor}$ at 95\%
confidence presented in column 7 of Table~IV as a function of
separation. Comparing column 7 with column 5, we conclude that the
error in the theoretical forces due to uncertainties in
experimental separations is in fact the most important factor in the
determination of the total theoretical error at short separations.
This concludes the analysis of the theoretical errors.

Using Eqs.~(\ref{eq15}) and (\ref{eq23}), the Casimir forces
$F^{\rm theor}(z_i)$ were computed at all experimental points for the
sample of resistivity $\rho$ used in experiment. A few results at 
different separations are presented in column 3 of Table~III. Column 5
of Table~III contains typical differences
$F^{\rm theor}(z_i)-{\bar{F}}^{\rm expt}(z_i)$ at different $z_i$.
The same computation was repeated for dielectric Si resulting
in Casimir forces ${\tilde{F}}^{\rm theor}(z_i)$ presented in
column 4 of Table~III. The typical differences
${\tilde{F}}^{\rm theor}(z_i)-{\bar{F}}^{\rm expt}(z_i)$ are
listed in column 7 of the same table. The comparison of both
sets of theoretical computations with experiment using a rigorous
statistical approach is performed in the next section.

\section{Comparison of experiment and theory as a test for
conductivity properties}

To compare experiment with theory, we consider the random quantity
$F_R^{\rm theor}(z_i)-{{F}}^{\rm expt}(z_i)$, where
$F_R^{\rm theor}$ is a stochastic function with a mean
${\bar{F}}_R^{\rm theor}=F^{\rm theor}$. In fact, in
Sec.~V we have discussed the uncertainties of just this
stochastic function $F_R^{\rm theor}$ 
and not of the usual function $F^{\rm theor}$
computed by the Lifshitz formula (\ref{eq15}) and corrected for the
presence of roughness in Eq.~(\ref{eq23}). Thus, it is implied
that 
$\Delta^{\!\rm tot}F_R^{\rm theor}=\Delta^{\!\rm tot}F^{\rm theor}$.
The absolute error of $F_R^{\rm theor}(z_i)-{{F}}^{\rm expt}(z_i)$
at 95\% confidence is denoted by $\Xi_{0.95}$. It can be found
using the composition rule (\ref{eq11}) with $J=2$ and
$k_{0.95}^{(2)}=1.1$:
\begin{equation}
\Xi_{0.95}(z_i)=\min\left\{
\Delta^{\!\rm tot}F^{\rm theor}(z_i)+
\Delta^{\!\rm tot}F^{\rm expt}(z_i),
1.1\sqrt{\left[\Delta^{\!\rm tot}F^{\rm theor}(z_i)\right]^2+
\left[\Delta^{\!\rm tot}F^{\rm expt}(z_i)\right]^2}\right\}.
\label{eq26}
\end{equation}
\noindent
Note that Eq.~(\ref{eq26}) provides us with the most conservative
estimate for this error because it is obtained under a supposition
that both $F_R^{\rm theor}(z_i)$ and ${{F}}^{\rm expt}(z_i)$ are
described by a uniform distribution. For other distribution laws
the absolute error $\Xi_{0.95}$ would be smaller. The confidence
interval for the mean value 
$F^{\rm theor}(z_i)-{\bar{F}}^{\rm expt}(z_i)$ of the random
quantity $F_R^{\rm theor}(z_i)-{{F}}^{\rm expt}(z_i)$ is
$[-\Xi_{0.95}(z_i),\Xi_{0.95}(z_i)]$. By the construction of this interval,
the mean $F^{\rm theor}(z_i)-{\bar{F}}^{\rm expt}(z_i)$ must
belong to it at 95\% confidence. The typical values of $\Xi_{0.95}$
at different separations are listed in column 6 of Table~III.

In the comparison of experiment with theory we will also need
the confidence interval $[-\Xi_{0.7},\Xi_{0.7}]$ obtained
at 70\% confidence. It is well known that for the normal
distribution
\begin{equation}
\frac{\Xi_{0.95}}{\Xi_{0.7}}=
\frac{t_{0.975}(\infty)}{t_{0.85}(\infty)}=2.
\label{eq27}
\end{equation}
\noindent
The distribution law of
$F_R^{\rm theor}(z_i)-{{F}}^{\rm expt}(z_i)$ can be investigated
by using the method for testing hypothesis about the form of the
distribution function of a random quantity \cite{28}.
As a result we find that the hypothesis of a normal distribution
is confirmed at all separations at probabilities larger
than 70\%. Note that for other distributions, distinct from
the normal, ${\Xi_{0.95}}/{\Xi_{0.7}}>2$ holds. Thus, by
putting below ${\Xi_{0.7}(z_i)}={\Xi_{0.95}}(z_i)/2$, we are in
fact conservative in the error analysis as the confidence interval
is wider than required.

Now we are in a position to compare experiment with theory.
In Fig.~5a we plot the differences 
$F^{\rm theor}(z_i)-{\bar{F}}^{\rm expt}(z_i)$ for all
experimental points over the separation range from 62.33 to
150\,nm where the total experimental error is less than 10\%.
The theoretical forces are computed as described in Sec.~V
for the Si sample of conductivity $\rho$ used in experiment.
The solid lines indicate the confidence interval
$[-\Xi_{0.95}(z_i),\Xi_{0.95}(z_i)]$. Dashed lines show
the confidence interval $[-\Xi_{0.7}(z_i),\Xi_{0.7}(z_i)]$.
As is seen from Fig.~5a, experiment and theory are consistent 
with the 95\% confidence interval. 
In fact, not only 95\% of individual points 
but all of them belong to the 95\% confidence interval.
What's more, not 30\% (as is permitted at 70\% confidence) but
only 10\% of all individual points are outside of the 70\%
confidence interval. This is a clear manifestation of the
fact that the theory is in excellent agreement with experiment
and that the above error analysis is very conservative
overestimating the above discussed errors and uncertainties.
The main reason for the overestimation is that we do not know
the exact magnitudes of the theoretical errors (such as due to
sample-to-sample variation of the optical data for Au and
dielectric Si, use of the proximity force theorem or
uncertainties in surface separations), and have therefore
replaced them with their upper limits in Sec.V.

In Fig.~5b the same information, as in Fig.~5a, is presented but
the differences 
${\tilde{F}}^{\rm theor}(z_i)-{\bar{F}}^{\rm expt}(z_i)$ 
are computed with the theoretical forces for dielectric Si.
As is seen from Fig.~5b, many points at all separations are
outside the 70\% confidence interval and practically all
of them at $z<100\,$nm. In Fig.~6,a,b we present the
differences ${{F}}^{\rm theor}-{\bar{F}}^{\rm expt}$
and ${\tilde{F}}^{\rm theor}-{\bar{F}}^{\rm expt}$, respectively,
on an enlarged scale within the separation range from 60 to
100\,nm. From Fig.~6a it is seen that the theory for the sample
of conductivity $\rho$, used in experiment, is consistent with
experimental data. There are no points outside of 95\%
confidence interval, and less than 3\% of all points are
outside of 70\% confidence interval (once again, this is the
indication that the errors were overestimated). A completely
different situation is observed in Fig.~6b. Here almost all
points representing the differences 
${\tilde{F}}^{\rm theor}(z_i)-{\bar{F}}^{\rm expt}(z_i)$ 
computed for dielectric Si (except of two) are outside the
70\% confidence interval. What this means is theory for the
dielectric Si is rejected by experiment at 70\% confidence
within the separation range from 60 to 100\,nm. The consistency
of the experimental data with the theoretical forces
${{F}}^{\rm theor}(z_i)$ and the rejection of the theory for
dielectric Si at 70\% confidence demonstrates the influence of
semiconductor conductivity on the Casimir force between a
metal and semiconductor.

This conclusion made using the rigorous statistical approach is
illustrated in a more usual way in Fig.~7 where the experimental
points are plotted with their error bars 
($\pm\Delta z,\pm\Delta^{\!\rm tot}F^{\rm expt}$) and the
theoretical dependences for the conductive and dielectric Si
are shown by the solid and dashed lines, respectively.
It is not possible to plot all experimental points with error
bars within a wide separation range from 60 to 100\,nm.
Because of this, to demonstrate that the result does not depend
on the selection of points, we plot points $1,11,21,\ldots$
in Fig.~7a and points $5,15,25,\ldots$ in Fig.~7b.
As is seen from Fig.~7,a,b, the solid lines are in very good
agreement with experiment, whereas the dashed lines deviate
significantly from experimental data. There are no noticeable
differences between figures obtained from different subsets of
data. For a more narrow range of separations from 75 to 90\,nm, 
in Fig.~8 we plot all experimental points with their error
bars and also theoretical lines for conductive (solid) and
dielectric (dashed) Si. Once again, the solid line is consistent
with experiment, whereas the dashed line is inconsistent.
It should be emphasized, however, that the results of the
rigorous statistical analysis, presented in Figs.~5 and 6, are
found at definite confidence probabilities and, thus, are more
informative than the qualitative results reflected on more
conventional figures 7 and 8.

\section{Conclusions and discussion}

In this paper we have continued the experimental investigation
of the Casimir force between a metal and a semiconductor which
had been started in Ref.~\cite{20}. The variety of electrical
properties inherent to semiconductors makes them attractive 
candidates for modifying the Casimir force by changing some of 
these properties, such as the resistivity. In Ref.~\cite{20} it was
only shown that in the Au-Si system the Casimir force decreases
more rapidly with separation than for Au-Au test bodies.
Here we demonstrate that the measurement of the Casimir force
can be used as an experimental test for semiconductor
conductivity properties. The measured force-distance relation 
for a Si plate of definite resistivity $\rho$ was compared with 
two theoretical values, first, for the sample, used in 
experiment, and, second, for a similar sample made of dielectric 
Si. As was demonstrated in Sec.~VI, the first computation is
consistent with data, whereas the second is rejected by the
experiment
at 70\% confidence. Thus, the measurement data for the Casimir
force between a metal and semiconductor are sensitive 
to semiconductor conductivity properties.

To achieve these results, the precision measurement of the Casimir
force was performed in high vacuum between specially prepared
and characterized Si plate and Au coated sphere. The total
relative experimental error of this measurement changes from
0.87 to 5.3\% in the separation range from 62.33 to 119.96\,nm.
Within the same separation range the total theoretical error 
decreases from 3.8\% to 2.0\%. The measure of agreement between
experiment and theory for the conductive semiconductor at
70\% confidence is of about 2\% within the separation range
from 62.33 to 100.07\,nm. Our conclusion that the
experimental data reject the hypothesis of the dielectric Si
was obtained at 70\% confidence. This is in fact strong evidence
considering the conservative character of the statistical 
analysis used.

The demonstrated dependence of the Casimir force between a metal 
and semiconductor on the conductivity properties of the latter can 
be applied to semiconductor microdevices. The density of charge
carriers in semiconductors can be changed either by doping and/or 
with time due to irradiation of the
semiconductor by laser light. The change of conductivity properties
results in the modulation of the Casimir force.
New experiments on modifying the Casimir
force between metal and semiconductor test bodies are planned using 
semiconductors of different conductivity due to different doping,
and  irradiation by laser light.

Of special interest is the measurement of the thermal effect in the
Casimir force both between metals and semiconductors. 
The present experiment is aimed at a better understanding of
the influence of semiconductor conductivity properties on the
Casimir force and does not have enough sensitivity to
measure the thermal corrections. 
The discussion of various proposed experiments on the
measurement of the thermal Casimir force at both short and large
separations can be found in Refs.~\cite{40,41,42,43,44,45}. 

\section*{Acknowledgment}
G.L.K. and V.M.M. are grateful to the Department of Physics of
the University of California (Riverside) for kind hospitality.
This work was supported by the NSF Grant PHY0355092 and
DOE grant DE-FG02-04ER46131. G.L.K. and V.M.M. were also partially
supported by Finep (Brazil).


\begingroup
\squeezetable
\begin{table}
\caption{Fractions $v_{k}$ of Au surface  covered by
roughness with heights $h_{k}$.}
\begin{ruledtabular}
\begin{tabular}{lcc}
$k$ & $h_k\,$(nm) & $v_k$ \\
\hline 
1&0&$8\times 10^{-5}$ \\
2&1&$8.5\times 10^{-4}$ \\
3&2&$1.21\times 10^{-3}$ \\
4&3&$1.6\times 10^{-3}$ \\
5&4&$4.09\times 10^{-3}$ \\
6&5&$4.77\times 10^{-3}$ \\
7&6&$4.71\times 10^{-3}$ \\
8&7&$5.62\times 10^{-3}$ \\
9&8&$1.111\times 10^{-2}$ \\
10&9&$1.671\times 10^{-2}$ \\
11&10&$2.591\times 10^{-2}$ \\
12&11&$4.148\times 10^{-2}$\\
13&12&$6.052\times 10^{-2}$\\
14&13&$8.644\times 10^{-2}$\\
15&14&$8.165\times 10^{-2}$\\
16&15&0.15265\\
17&16&0.1262\\
18&17&0.107\\
19&18&$9.802\times 10^{-2}$\\
20&19&$6.958\times 10^{-2}$\\
21&20&$4.98\times 10^{-2}$\\
22&21&$2.58\times 10^{-2}$\\
23&22&$1.288\times 10^{-2}$\\
24&23&$6.4\times 10^{-3}$\\
25&24&$3.29\times 10^{-3}$\\
26&25&$1.11\times 10^{-3}$
\end{tabular}
\end{ruledtabular}
\end{table}
\endgroup
\begingroup
\squeezetable
\begin{table}
\caption{Fractions $v_{l}$ of Si surface covered by
roughness with heights $h_{l}$.}
\begin{ruledtabular}
\begin{tabular}{lcc}
$l$ &$h_l\,$(nm)& $v_l$ \\
\hline 
1&0&$2.0\times 10^{-5}$ \\
2&0.1&$8.1\times 10^{-4}$ \\
3&0.2&$8.84\times 10^{-3}$ \\
4&0.3&$4.27\times 10^{-2}$ \\
5&0.4&0.10384 \\
6&0.5&0.34379 \\
7&0.6&0.3683 \\
8&0.7&$9.9\times 10^{-2}$ \\
9&0.8&$3.05\times 10^{-2}$ \\
10&0.9&$2.13\times 10^{-3}$ \\
11&1.0&$7.0\times 10^{-5}$ 
\end{tabular}
\end{ruledtabular}
\end{table}
\endgroup
\begingroup
\squeezetable
\begin{table}
\caption{Mean experimental (${\bar{F}}^{\rm expt}$) and
theoretical (${{F}}^{\rm theor},\,{\tilde{F}}^{\rm theor}$)
Casimir forces computed for different Si samples,
differences of these forces and their confidence intervals
($\Xi_{0.95},\,\Xi_{0.7}$) at confidence probabilities 95
and 70\% in dependence on separation (see text for details).
}
\begin{ruledtabular}
\begin{tabular}{cccccccc}
$z_i$&${\bar{F}}^{\rm expt}$&${{F}}^{\rm theor}$&
${\tilde{F}}^{\rm theor}$&${{F}}^{\rm theor}-{\bar{F}}^{\rm expt}$&
$\Xi_{0.95}$&${\tilde{F}}^{\rm theor}-{\bar{F}}^{\rm expt}$&
$\Xi_{0.70}$\\
(nm)&(pN)&(pN)&(pN)&(pN)&(pN)&(pN)&(pN)\\
\hline 
62.33&--380.0&--380.5&--374.4&--0.50&15.2&5.7&7.6\\
69.98&--280.9&--277.9&--272.9&3.0&10.4&8.0&5.2\\
80.01&--196.4&--192.8&--188.9&3.6&7.1&7.5&3.55\\
90.04&--140.4&--139.4&--136.3&1.0&5.4&4.1&2.7\\
100.07&--106.2&--104.2&--101.7&2.0&4.5&4.5&2.25\\
109.93&--80.30&--80.35&--78.23&--0.05&4.1&2.1&2.05\\
119.96&--62.90&--63.05&--61.26&--0.15&3.9&1.64&1.95\\
140.02&--40.98&--40.96&--39.64&0.02&3.8&1.35&1.9\\
160.08&--26.93&--28.14&--27.11&--1.2&3.7&--0.19&1.8\\
180.14&--19.70&--20.18&--19.36&--0.48&3.7&0.34&1.8\\
200.03&--14.71&--15.02&--14.35&--0.31&3.7&0.36&1.8\\
250.18&--7.132&--7.968&--7.539&--0.84&3.7&--0.41&1.8\\
299.99&--5.221&--4.756&--4.455&0.46&3.7&0.76&1.8
\end{tabular}
\end{ruledtabular}
\end{table}
\endgroup
\begingroup
\squeezetable
\begin{table}
\caption{Different experimental and theoretical relative errors 
(\%) in dependence on separation (see text for details).
}
\begin{ruledtabular}
\begin{tabular}{lcccccc}
$z_i\,$(nm)&$\delta^{\rm rand}F^{\rm expt}$ &
$\delta^{\rm syst}F^{\rm expt}$&$\delta^{\rm tot}F^{\rm expt}$&
$\delta_0F^{\rm theor}$&$\delta_3F^{\rm theor}$&
$\delta^{\rm tot}F^{\rm theor}$\\
\hline 
62.33&0.78&0.31&0.87&0.55&3.8&3.8\\
69.98&1.1&0.42&1.2&0.56&3.4&3.4\\
80.01&1.6&0.60&1.7&0.56&2.9&2.9\\
90.04&2.1&0.84&2.4&0.56&2.7&2.7\\
100.07&2.9&1.1&3.1&0.56&2.4&2.4\\
109.93& 3.7&1.4&4.1&0.56&2.2&2.2\\
119.96&4.7&1.8&5.3&0.56&2.0&2.0\\
140.02&7.3&2.8&8.1&0.57&1.8&1.9\\
160.08&10&4.1&12&0.58&1.5&1.7\\
180.14&15&5.7&17&0.58&1.4&1.6\\
200.03&20&7.7&22&0.59&1.2&1.4\\
250.18&42&16&47&0.61&1.0&1.3\\
299.99&57&22&64&0.64&0.9&1.2
\end{tabular}
\end{ruledtabular}
\end{table}
\endgroup

\begin{figure}
\vspace*{-7cm}
\centerline{
\includegraphics{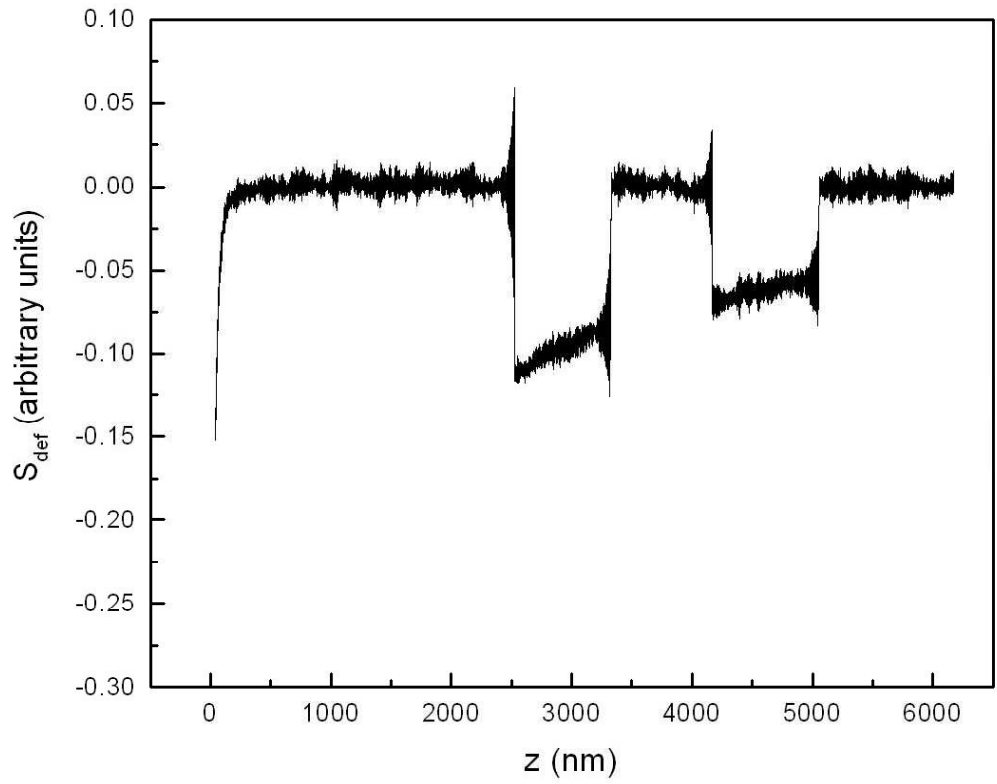}
}
\vspace*{-9cm}
\caption{The deflection signal of the cantilever in responce to the
dc voltage and two square voltage pulses applied to the Si plate
as a function of separation.}
\end{figure}
\begin{figure}
\vspace*{-7cm}
\centerline{
\includegraphics{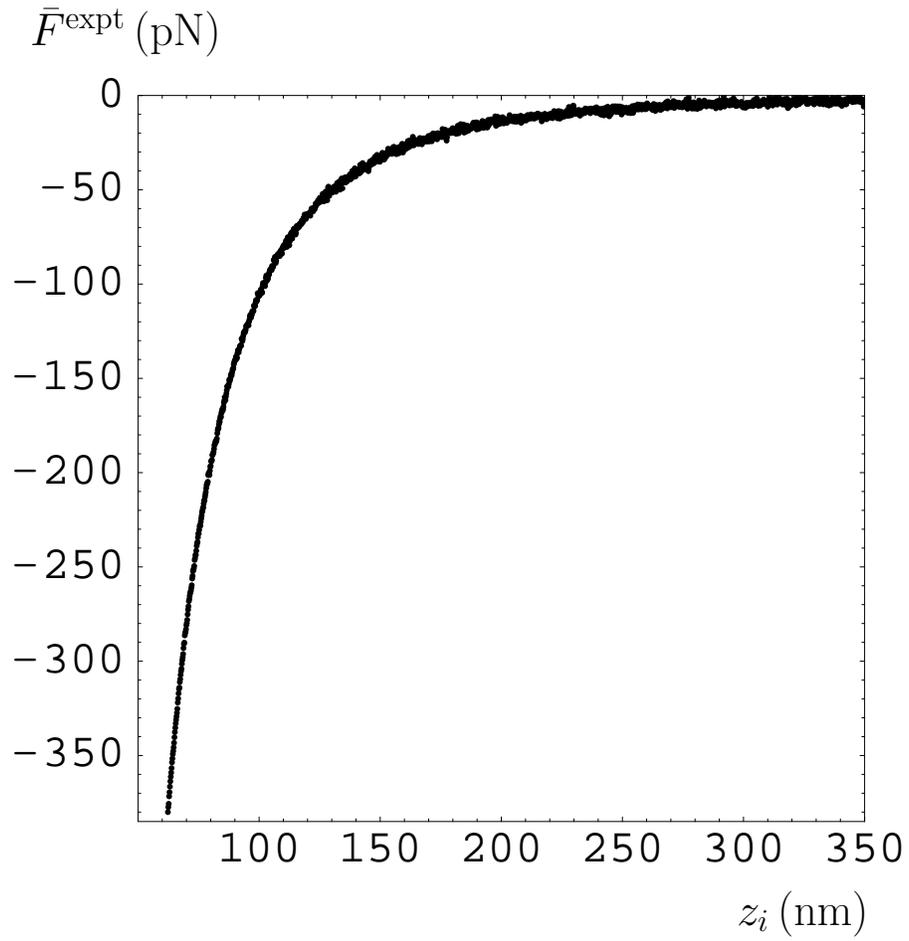}
}
\vspace*{-6cm}
\caption{The mean measured Casimir
force  as a function of separation between 
the zero roughness levels of Si plate and Au sphere. }
\end{figure}
\begin{figure}
\vspace*{-7cm}
\centerline{
\includegraphics{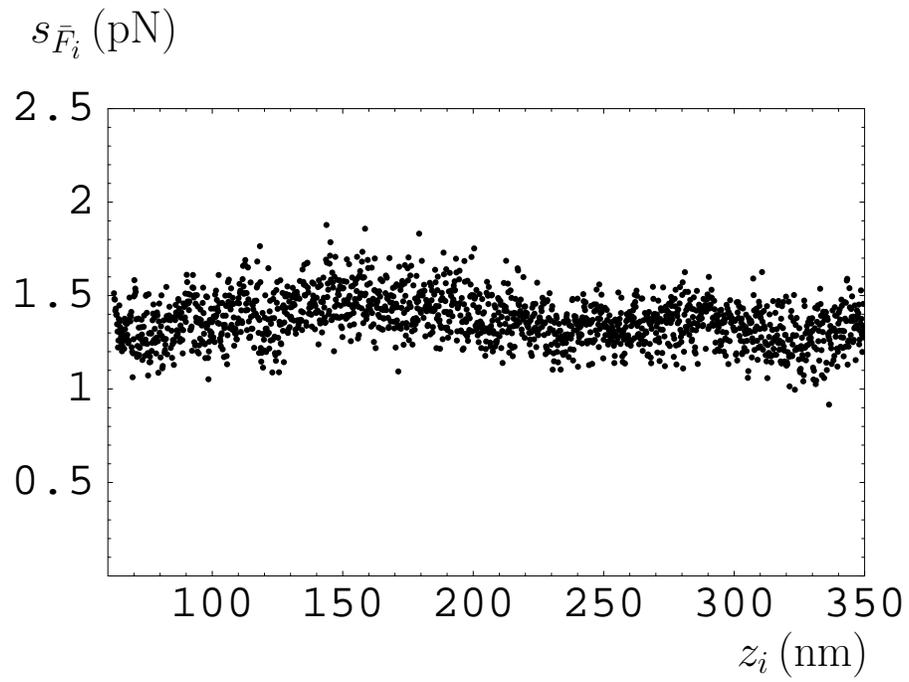}
}
\vspace*{-6cm}
\caption{The estimate for the variance of the mean measured 
Casimir force  as a function of plate-sphere separation.}
\end{figure}
\begin{figure}
\vspace*{-7cm}
\centerline{
\includegraphics{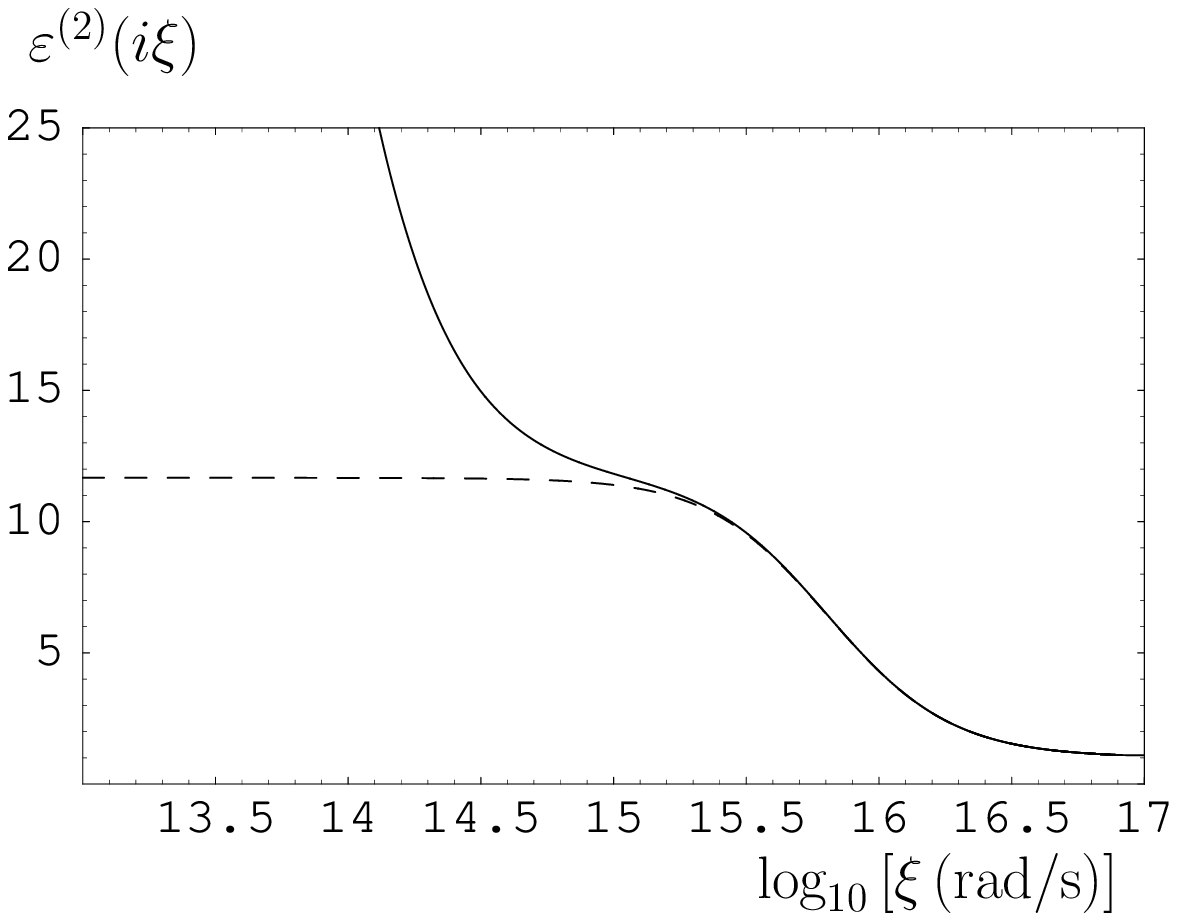}
}
\vspace*{-6cm}
\caption{
Dielectric permittivity of Si plate used in experiment
along the imaginary frequency axis (solid line).
Dashed line shows the dielectric permittivity
${\tilde{\varepsilon}}^{(2)}(i\xi)$ of a dielectric Si.
}
\begin{figure}
\vspace*{-1cm}
\centerline{
\includegraphics{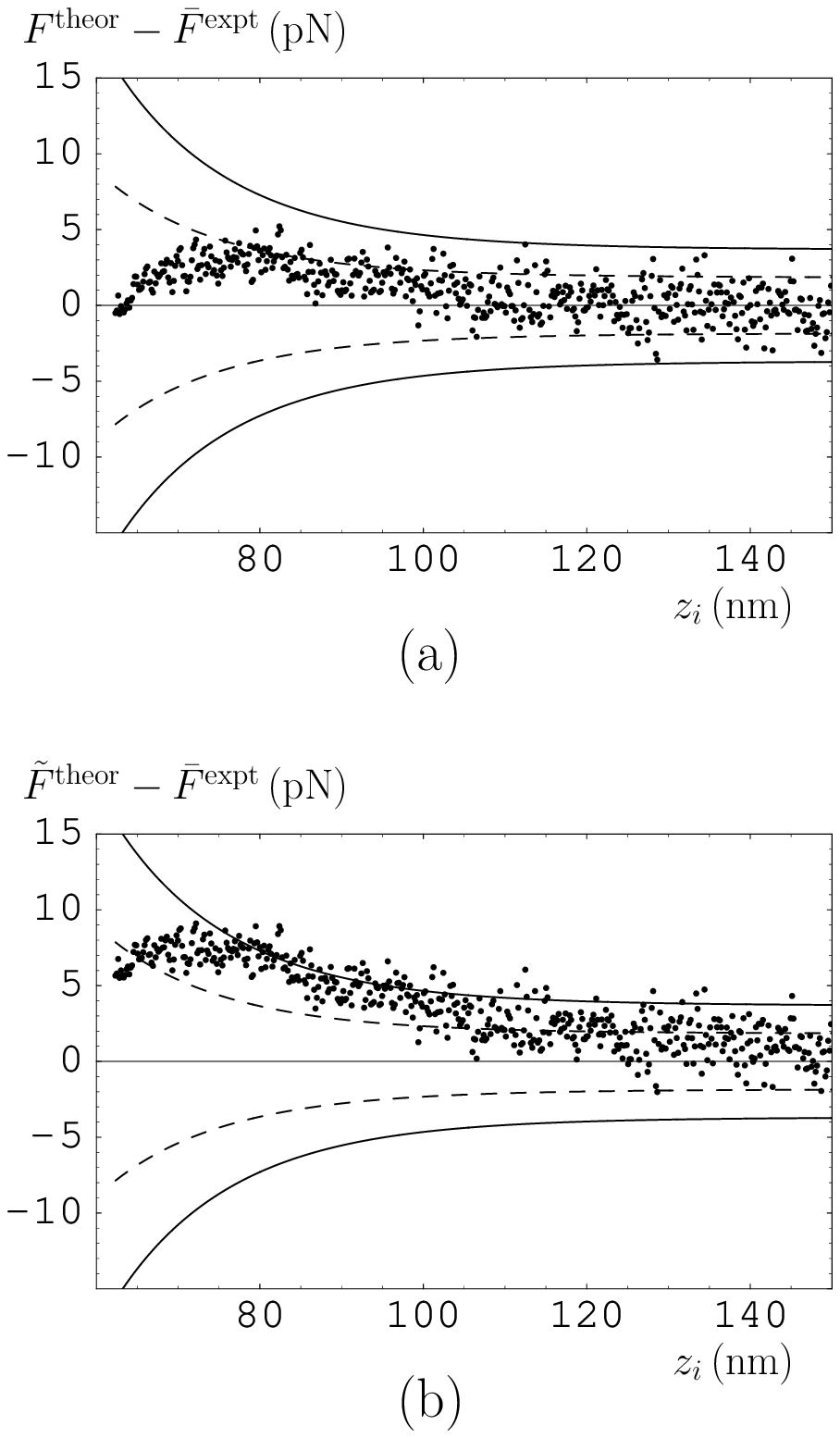}
}
\vspace*{-11cm}
\caption{
Differences of the theoretical and mean experimental Casimir
forces versus separation. Theoretical forces are computed for 
(a) the
Si plate used in experiment and (b) for dielectric Si.
Solid and dashed lines indicate 95 and 70\% confidence
intervals, respectively.
}
\end{figure}
\begin{figure}
\vspace*{-1cm}
\centerline{
\includegraphics{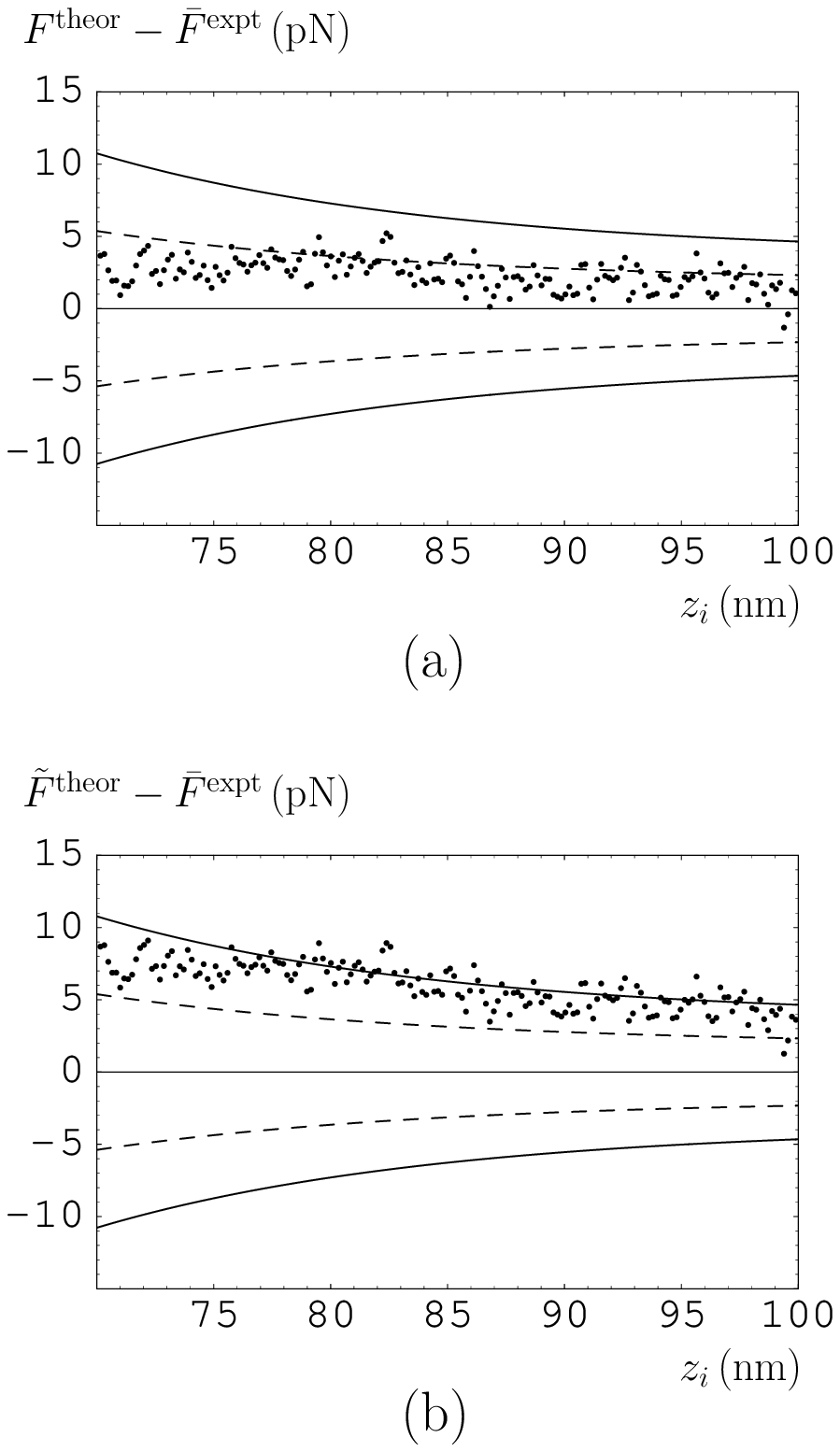}
}
\vspace*{-11cm}
\caption{
Differences of the theoretical and mean experimental Casimir
forces versus separation plotted on an enlarged scale. 
Theoretical forces are computed (a) for the
Si plate used in experiment and (b) for dielectric Si.
Solid and dashed lines indicate 95 and 70\% confidence
intervals, respectively.
}
\end{figure}
\begin{figure}
\vspace*{-4cm}
\centerline{
\includegraphics{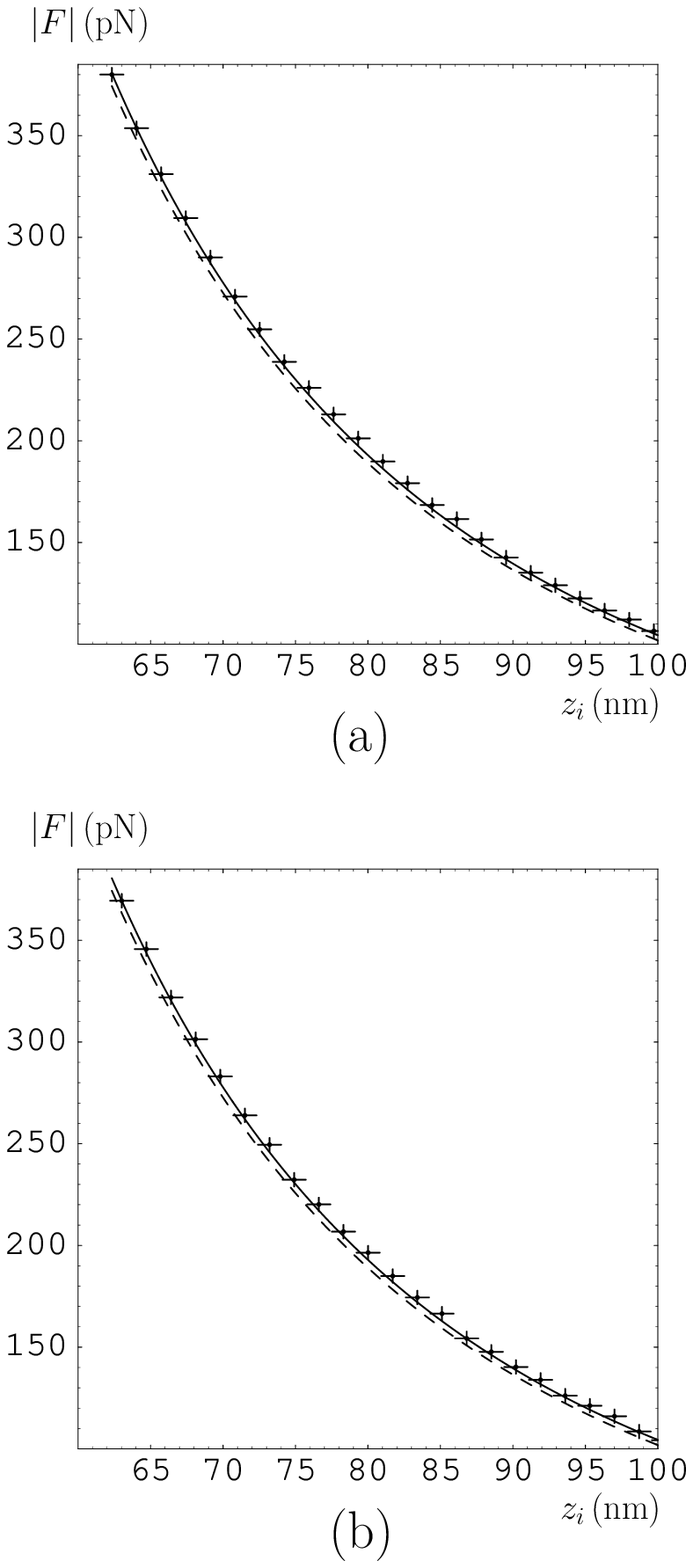}
}
\vspace*{-7.5cm}
\caption{
The magnitudes of the experimental Casimir forces with their error 
bars versus separation for the points (a) $1,11,21,\ldots$ and
(b) $5,15,25,\ldots$. Solid lines show the theoretical dependence
for the sample used in experiment and dashed lines for the
dielectric Si.
}
\end{figure}
\begin{figure}
\vspace*{-6cm}
\centerline{
\includegraphics{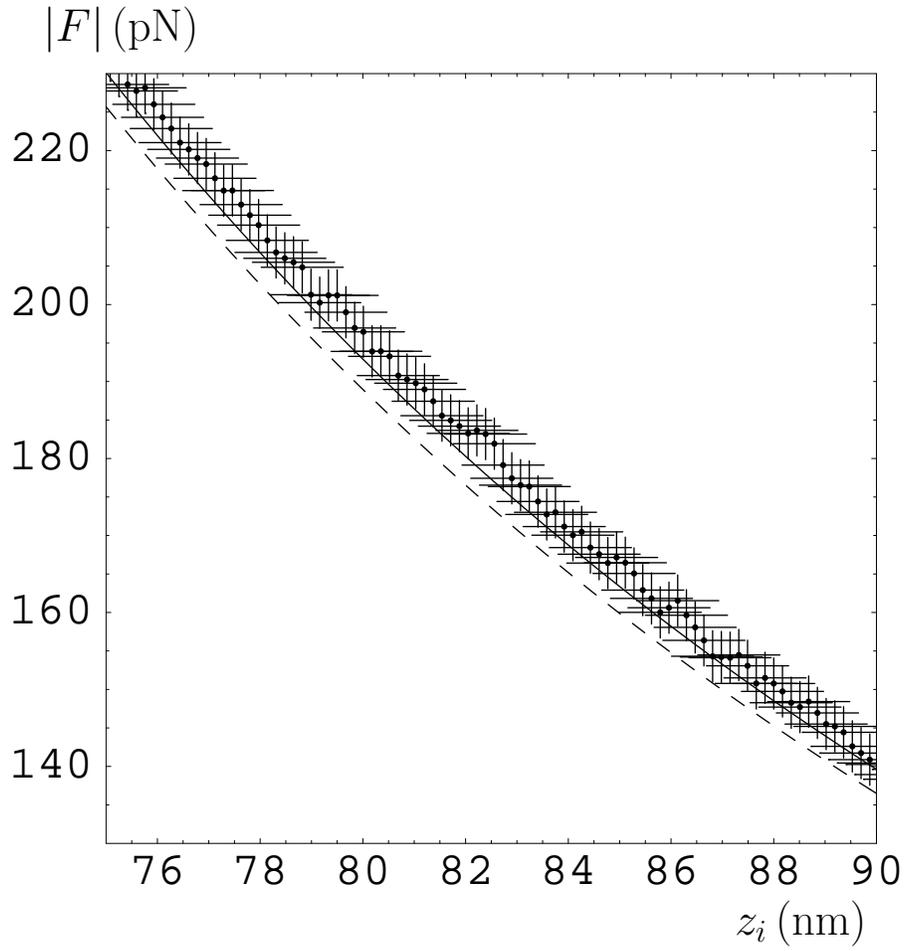}
}
\vspace*{-9cm}
\caption{
The magnitudes of the experimental Casimir forces with their error 
bars versus separation.
Solid line shows the theoretical dependence
for the sample used in experiment and dashed line for the
dielectric Si.
}
\end{figure}
\end{document}